 \def \half {{1\over 2}}
\newcommand{\sss}{{\scriptscriptstyle }}

\def\eV{\hbox{eV}}
\def\TeV{\hbox{TeV}}
\def\roughly#1{\mathrel{\raise.3ex\hbox{$#1$\kern-.75em
\lower1ex\hbox{$\sim$}}}}
\def\lsim{\roughly<}
\def\gsim{\roughly>}
\def\pref#1{(\ref{#1})}

\documentstyle[prl,aps]{revtex}
\begin{document}
\twocolumn[\hsize\textwidth\columnwidth\hsize\csname@twocolumnfalse%
\endcsname
\rightline{McGill-01/10, {\tt hep-th/0105261}}
\vspace{3mm}

\draft
\title{Exponentially Large Extra Dimensions}

\author{A. Albrecht,${}^a$ C.P.~Burgess,${}^{b,c}$ F. Ravndal ${}^d$ 
and C. Skordis ${}^a$}

\address{${}^a$ Department of Physics, UC Davis,
1 Shields Avenue, Davis CA, USA 95616.\\
${}^b$ School of Natural Sciences, Institute for Advanced Study, 
Princeton NJ, USA 08540.\\
${}^c$ Physics Department, McGill University,
3600 University Street, Montr\'eal, Qu\'ebec, Canada H3A 2T8.\\
${}^d$ Institute of Physics, University of Oslo, N-0316 Oslo, Norway.}
\maketitle

\begin{abstract}
{
We show how the presence of a very light scalar with a cubic self-interaction 
in six dimensions can stabilize the extra dimensions at radii which are naturally exponentially 
large, $r \sim \ell \exp [(4\pi)^3/g^2]$, where $\ell$ is a microscopic physics scale
and $g$ is the (dimensionless) cubic coupling constant. The resulting radion mode of
the metric becomes a very light degree of freedom whose mass, $m \sim 1/(M_p r^2)$ is stable under
radiative corrections. For $1/r \sim 10^{-3}$ eV the radion is extremely light, 
$m \sim 10^{-33}$ eV. Its couplings cause important
deviations from General Relativity in the very early universe, but naturally evolve
to phenomenologically acceptable values at present. 
}
\end{abstract}
\pacs{PACS numbers: }
]

%
%

\section{Introduction}

Recent developments in both particle physics and cosmology appear to indicate
the existence of important fundamental physics associated with the scale 
$10^{-3}$ eV. 

\begin{itemize}
\item
On the cosmological side, physics at this scale appears to
be indicated by the recent observational evidence \cite{Cosmology}
for the existence of a `dark energy' component to the universe 
(possibly a cosmological constant), having a negative pressure of 
order $p \sim (10^{-3} \, \eV)^4$. 

\item
On the particle physics side several developments have made plausible
the existence of interesting $10^{-3}$ eV physics. On the one hand, it 
has long been recognized that breaking supersymmetry at the TeV scale 
would imply masses of order $m \sim (1 \TeV)^2/M_p \sim 10^{-3}$ eV
for the gravitino, and other gravitationally-coupled particles. Here
$M_p \sim 10^{18}$ GeV is the (rationalized) Planck mass.

\item
More recently has come the recognition that extra dimensions can be much 
larger (and have much richer dynamics) than had hitherto been appreciated,
and the realization that such large extra dimensions could help
solve some long-standing problems like the hierarchy problem \cite{ADD,RS}.  
In particular, these dimensions could have radii as large as 
$r \sim (10^{-3} \eV)^{-1}$. 
\end{itemize}

Interest in all of these issues has been sharpened by the realization that
the existence of new particles at this scale may have other observational
consequences. The resulting modifications to Newton's Law of Gravity may
fall within reach of current and upcoming experiments. The dynamics of
red giants and supernovae can be modified, and the resulting bounds can
be important \cite{SNProbs,LED}, but need not be fatal \cite{SNnotfatal,OtherLogs}.

A crucial part of any large-extra-dimension scenario is a natural
mechanism for generating a radius which is large compared to other
microphysical scales. This is a particularly pointed requirement if
the large ratio, $(10^{-3} \eV)/(1 \TeV)$, is to be used to explain 
other hierarchies, like $M_{\sss W}/M_p$. 

It is our purpose in this paper to propose a mechanism for generating
such large dimensions. Although our proposal is not restricted to 
radii as large as $r \sim (10^{-3} \eV)^{-1}$, it can satisfy the
very restrictive phenomenological constraints which apply in this
case. Furthermore, we argue that the energetics which
chooses the value for the extra-dimensional radius in our mechanism
can have attractive cosmological consequences when the extra dimensions
are large. 

Our proposal is based on the observation that
large radii are naturally obtainable if the potential which governs 
the radion is logarithmic:\footnote{For another approach to obtaining 
logarithmic radion potentials in six dimensions, see \cite{OtherLogs}.}
\begin{eqnarray}
\label{LogPot} 
\ell^4 V(r) &=& \left({\ell\over r}\right)^p 
\left[ a_0 + a_1 \log \left({r\over \ell}\right)  + \cdots \right] \nonumber \\
&+& {\cal O}\left[\left({\ell\over r}\right)^{q}\right],
\end{eqnarray}
(where $q>p$ and $\ell$ is a microscopic length scale). Besides the usual
runaway solution, $r \to \infty$, the stationary condition
$dV/dr = 0$ also admits the solutions:
\begin{equation}
\label{LogPotSoln}
{r \over \ell} \approx \exp\left({1\over p} - {a_0 \over a_1} \right) ,
\end{equation}
where the approximation becomes exact in the absence of higher powers
of $\log(r/\ell)$.  This basic mechanism was proposed at a
phenomenological level in the ``Planck scale quintessence''
models\cite{AS1}. In that context it was shown that the time for 
quantum tunneling out of the local minimum was exponentially larger
than the age of the Universe \cite{WellerBubble}.

Eq.~\pref{LogPotSoln} predicts $r$ is naturally exponentially large 
compared to $\ell$, if two conditions are satisfied:
\begin{enumerate}
\item
$a_0$ and $a_1$ must have opposite signs; and 
\item
there is a modest hierarchy in the coefficients, $a_k$. 
\end{enumerate}
For instance if $a_1 = - \hat{a}_1 \epsilon$, with $\epsilon <1$ and
$a_0$ and $\hat{a}_1$ both positive and ${\cal O}(1)$, then $r/\ell  = {\cal O}(\exp[a_0/
(\hat{a}_1 \epsilon)+{\cal O}(1)]) \gg 1$. Numerically, if $1/\ell \sim 1$ TeV
and $a_0/(\hat{a}_1 \epsilon) \sim 35$ then $1/r \sim 10^{-3}$ eV
falls into the interesting range. 

We here propose a scenario which generates logarithmic radion potentials 
which very generically satisfy both of these two conditions. The scenario 
has the radion potential generated as the universe passes through a 
stage during which it is effectively six-dimensional, provided
that there is at least one six-dimensional scalar field whose mass is 
of order $1/r$ and which has reasonably large, nonderivative,
cubic self-interactions. 

In six dimensions a cubic scalar self-interaction, 
\begin{equation}
\label{CubicInt}
U_{\rm ren}(\phi) = {g\over 3!} \; \phi^3 ,
\end{equation}
is the only local interaction which has a dimensionless coupling constant. 
Because of this, in perturbation theory its renormalization gives it
a logarithmic, rather than a power, dependence on $r$. In 
renormalization-group terms it is the only six-dimensional interaction which 
is not irrelevant in the infrared. 

The logarithmic dependence of the low-energy coupling, $g(r)$, provides
a natural way of obtaining logarithmic potentials for $r$. In the absence
of light-scalar loops, for large $r$ the radion potential has the
generic form 
\begin{equation}
\label{GenForm}
V(r) = \sum_{k=k_0}^\infty {c_k \over r^k}.
\end{equation}
This can arise
in particular examples in many ways. For instance, it arises when evaluating
the classical action as a function of radius if six-dimensional gravity 
or supergravity is compactified on a sphere. Alternatively, it could 
dominantly arise as a quantum Casimir energy, such as in a toroidal 
compactification \cite{PontonPoppitz}.

Scalar radiative corrections (in six dimensions)
to this potential correct the constants $c_k$:
\begin{equation}
\label{SeriesForm}
c_k = \sum_{l=l_0}^\infty  c_k^{(l)} \, \left[ {g^2(r) \over (4\pi)^3} \right]^l,
\end{equation}
where $g(r) = g_0 + b g_0^3 \log(r/\ell) + {\cal O}(g_0^5)$. 

Using eq.~\pref{SeriesForm} in eq.~\pref{GenForm} produces a logarithmic
potential of the desired type, with several remarkable features:
\begin{itemize}
\item
Eq.~\pref{SeriesForm} automatically introduces the desired hierarchy (and so
satisfies condition 2 above) by
systematically suppressing higher powers of $\log r$ with the suppression factor 
$\epsilon \sim \alpha = g^2/(4\pi)^3$. 
\item
The relative sign of the coefficients, $a_0$ and $a_1$, of the first two
terms in eq.~\pref{LogPot} depends crucially on the relative sign of the
first few loop corrections -- $c_{k_0}^{(l_0)}$ and $c_{k_0}^{(l_0+1)}$ -- and
on the sign of $b$, the one-loop renormalization-group coefficient for $g$. 
Furthermore, given that the first term in $V$ is positive ($c_{k_0}^{(l_0)}>0$,
the signs of $a_o$ and $a_1$ are opposite, as required to generate a hierarchy 
(condition 1 above). 
\item
Because the dependence of $V(r)$ on $r$ arises implicitly through the dependence
of $V$ on $\alpha(r)$, the extremal value for $r=r_s$ corresponds to the coupling
$\alpha_s = \alpha(r_s)$ which extremizes $V(r)$. Generically, if all constants
$c_{k_0}^{(l)}$ are ${\cal O}(1)$, the minimum occurs when $\alpha(r_s) = {\cal O}(1)$, and so
the precise value for $r=r_s$ at the minimum cannot be computed perturbatively in 
$\alpha(r_s)$. The hierarchy $r/\ell$ is nonetheless reliably predicted to 
be exponentially large, so long as $\alpha(r_s)$ differs significantly from
$\alpha(\ell)$ (and {\it either} $\alpha(\ell)$ or $\alpha(r_s)$ lies within
the perturbative regime) given the logarithmic running of $\alpha$. 

Perturbative calculation of $r_s$ itself is also possible if the lowest-order 
coefficient, $c^{\left(l_0\right)}_{k_0}$, happens to be even modestly 
smaller than the others (as occurs in some examples below).  
\end{itemize}

We present our argument in more detail in the subsequent sections. First, the
next section gives more explicit expressions for the radion potential in a 
model consisting only of the metric and a very light scalar field. Since our
mechanism expresses the explanation for large dimensions in terms of a
light scalar field, we then follow with a discussion of how natural it is
to find scalars with the required properties. We conclude with a brief
discussion of the cosmological and phenomenological implications of the 
logarithmic potential, and the bounds which these may impose on model building.

\section{Radion Potentials and Light Scalars in 6 Dimensions}

In this section we compute explicitly the radion potential produced
by a light scalar field for a simple compactification. To this end
consider a model consisting of scalar fields, $\varphi_i$, and the
six-dimensional metric, ${\cal G}_{\sss MN}$. We imagine this to be an effective
six-dimensional theory obtained after integrating out all more massive
degrees of freedom at scale $\ell$. The leading terms in the derivative
expansion for this lagrangian have the form:
\begin{equation}
\label{Lagr}
{{\cal L} \over \sqrt{{\cal G}}} = - \,{1\over 2 \ell^4 } {\cal R} - \half \,
{\cal G}^{\sss MN} \, \partial_{\sss M} \varphi_i \partial_{\sss N} \varphi_i
-U(\varphi),
\end{equation}
where ${\cal R}$ denotes the scalar curvature built from the six-dimensional
metric.

We assume the scalar potential to have the form
\begin{eqnarray}
\label{PotAssn}
U(\varphi) &=& { \mu^2_{ij}\over 2}\;  \varphi_i \varphi_j +{g_{ijk} \over 3!} \; \varphi_i
\varphi_j \varphi_k \\
&& \qquad  + {\kappa_{ijkl} \ell^2\over 4!}
\; \varphi_i \varphi_j \varphi_k \varphi_l + \cdots
\end{eqnarray}
with the microscopic scale $\ell$ setting the dimensions of all but two of the
couplings in the scalar potential. The two exceptions are: $(i)$ we assume
there is no cosmological constant term in $U(\varphi)$; and $(ii)$ 
we assume the scalar masses are small: $\mu_{ij} \lsim 1/r \ll 1/\ell$. In this section
we simply fine-tune the lagrangian to ensure these conditions are satified, but
since our generation of the logarithmic potential is based on these choices, in the
next section we address how difficult they are to arrange within models. Although
we shall argue that assumption $(ii)$ is simple to arrange in supersymmetric
models, the tricky part is to have $\mu$ be as small as ${\cal O}(1/r)$ without also
finding the cubic term similarly suppressed, $g_{ijk} \sim \ell^2/r^2$.

We consider in detail the two simplest cases. The first is a single real scalar, $\varphi$, 
with self-coupling $U = {1 \over 2}\mu^2\varphi^2 + {1\over 3!} g\varphi^3 + 
{1\over 4!} \kappa \ell^2 \varphi^4 + \dots$. The second is a single complex scalar,
$\phi = (\varphi_1 + i \varphi_2)/\sqrt{2}$, with $Z_3$ symmetry $\phi \to \omega \phi$,
where $\omega^3=1$. The self coupling in this case is  
$U =  \mu^2|\phi|^2 + \left( {1\over 6} g\phi^3 + c.c.\right)
+ {1\over 4} \kappa \ell^2 |\phi|^4 + \dots$.

\subsection{One Loop Casimir Energy}

Before searching for logarithmic corrections due to the cubic scalar coupling, 
we must first compute the potential of the form eq.~\pref{GenForm} which is to 
be corrected. This we compute in a semiclassical expansion about the local
minimum at $\phi = 0$, whose existence is assured by our choice
$\mu^2 >0$. Although this minimum is ultimately destabilized
by the assumed cubic term, we assume the potential to be bounded below by virtue 
of the other terms in $U(\phi)$, involving higher powers of $\phi$.
The detailed form of these higher terms do not play a role in the 
discussion which follows.

Our calculation also assumes the ground state geometry is flat: 
$M^6 = R^4 \times T^2$, where $R^4$ denotes flat Minkowski space and $T^2$ 
is a torus, both of whose radii we denote by $r$. (Although the torus has 
other moduli besides its radius, here we focus only on $r$.) For the
complex field, $\phi$, we allow the possibility that the scalars satisfy 
twisted boundary conditions -- $\phi \to \omega \phi$, with 
$\omega^3 = 1$ -- about the cycles of the torus. Indeed this is our 
primary reason for considering the complex scalar case.

Under these circumstances the radion potential first arises as a one-loop 
Casimir energy for the fields living in the bulk. (The Casimir energy for
fields on the brane do not depend directly on $r$ in the absence of a
nontrivial `warp factor', such as we take to be the case here.)
The contribution to this 
energy from a complex scalar is computed as
a function of its boundary conditions in Appendix A,
and is given by:
\begin{eqnarray}
\label{OneLoopResult}
V_1(r) &=& -\; {1 \over r^4} \int_0^\infty {dx \over x^3}\;
e^{-\beta x}\;\left[e^{-\pi x(a^2 + b^2)}\right. \nonumber \\
&\times & \left.\theta_3(i\pi a x,e^{-\pi x}) \,\theta_3(i\pi b x,
e^{-\pi x}) - {1\over x} \right], 
\end{eqnarray}
where $\beta = \mu^2 r^2 /(4 \pi)$, and $\theta_3(z,q)$ denotes the usual 
Jacobi theta-function \cite{WW}. The constants $a$ and $b$ take the
values $0,1/3$ or $2/3$ depending on the type of twisted boundary condition which 
the scalar satisfies about each of the torus' two nontrivial cycles. 
The choice $a,b=0$ corresponds to periodic boundary conditions about these
cycles, while $a,b=1/3,2/3$ corresponds to twisting by $\omega$ or $\omega^2$. 

For real scalars no twist consistent with a cubic self-interaction is possible,
and so the result is one half the expression of eq.~\pref{OneLoopResult}
with $a=b=0$.

As is easily verified (see Appendix A), eq.~\pref{OneLoopResult} 
converges in both the ultraviolet
and infrared, even if $\mu \to 0$. If $\mu r \gg 1$ then $V_1$ falls exponentially
as $\mu r \to \infty$. If $\mu \to 0$, then the potential takes the form of 
a power of $r$: $V_1 = c_4^{(1)}/r^4$ where the coefficient $c_4^{(1)}$ is given
by the integral in eq.~\pref{OneLoopResult} with $\beta = 0$.
Numerical integration gives the results shown in Table (1) below.

\smallskip
\begin{center}
\begin{tabular}{|c||r|r|}
\hline 
 ($a,\,b$)  &   $c_4^{(1)}$  &  $c_4^{(2)}$   \\
\hline 
 (0,\,0)       & $- 0.299 $  & $- 0.064$     \\
 (0,\,1/3)     & $- 0.048  $ &   0.178     \\
 (0,\,2/3)     & $- 0.048  $ &   0.327     \\
 (1/3,\,1/3)   &   0.122   &   0.069     \\
 (1/3,\,2/3)   &   0.122   &   0.142     \\
 (2/3,\,2/3) &   0.122   &   0.169     \\
\hline
\end{tabular}
\end{center}
\medskip
{\small Table (1): The one- and two-loop Casimir-energy coefficients, $c_4^{(1)}$
and $c_4^{(2)}$.}
\smallskip

To this should be added the contribution due to the graviton Casimir energy,
as well as the Casimir energy due to any other six-dimensional particles. Because
of the rapid falloff in the result as $\mu r \to \infty$ it is clear that only
the contribution of those degrees of freedom for which $\mu r \lsim 1$ is important
for large $r$.
We consider these contributions in more detail in following sections, where the
structure of the entire model is considered in more detail.

\subsection{Radiative Corrections}

We next turn to the corrections to $V_1(r)$ which dominate for large $r$. 


The first observation to be made is that only the cubic interaction of $U(\varphi)$ 
can contribute to $V(r)$ unsuppressed by further powers of $1/r$. This is because
all other coupling constants have dimension of a positive power of length, and
so are perturbatively nonrenormalizable. 

To see how this works consider the potential $U(\varphi)$ for a real scalar. 
Imagine now scaling out powers of $r$ to make all couplings dimensionless, 
so the potential $U(\varphi)$ is written
\begin{equation}
\label{RGPotForm}
U(\varphi) = {(\mu r)^2 \over r^2} \; \varphi^2 + {g \over 3!} \; \varphi^3 +
\left( {\kappa \ell^2 \over r^2} \right) \, r^2 \varphi^4 + \cdot
\end{equation}
Once written this way it is clear that on dimensional grounds the Casimir energy 
can be written: 
\begin{equation}
V =  {1\over r^4}v(\mu r, g, \kappa \ell^2/r^2,\dots) \; ,
\end{equation}
where $v$ is a dimensionless function of dimensionless arguments. Clearly
each factor of the coupling $\kappa$ in a series expansion of $v$ 
is accompanied by a power of $\ell^2/r^2$, implying a contribution which
is further suppressed by powers of $1/r$ compared to the uncorrected 
term \cite{IRDivs}.

For corrections to $V(r)$ which are {\it not} suppressed by more powers of $1/r$
we must consider graphs which involve only the dimensionless cubic scalar self-coupling
$g$. The emergence of the logarithms can then be most easily seen in the limit as
$\mu \to 0$, in which case it is revealed by a simple renormalization-group argument.
For $\mu \to 0$ the dominant radion corrections can be written, on dimensional grounds,
as:
\begin{equation}
\label{RGSeries}
V(r) = {1\over r^4} \Bigl[ A_0 +  A_1(r/r_0) \alpha(r_0) + A_2(r/r_0) \alpha^2(r_0)
+ \cdots \Bigr],
\end{equation}
where $\alpha = g^2/(4\pi)^3$ is the six-dimensional loop-counting parameter, 
renormalized at an arbitrary 
renormalization point, $r_0$. The coefficients $A_k(r/r_0)$ can be dimensionless
functions of $r/r_0$, although explicit calculation has
just shown $A_0 = c_4^{(1)}$ to be a constant. 

Since the dependence of $V$ on $r$ is tied to its dependence on $r_0$,
and since $V$ cannot depend on $r_0$ at all, the $r$ dependence of the $A_k$'s 
can be related to the running of $\alpha$. That is, if: 
\begin{equation}
\label{AlphaBetaFn}
r_0 {d\alpha \over dr_0} = B \alpha^2 + O(\alpha^3),
\end{equation}
then the Callan-Symanzik equation, $r_0 dV/dr_0 = 0$, implies $dA_1/dr = 0$ and
$r dA_2/dr = B\, A_1$. For renormalization schemes for which $B$ is $r$-independent, 
this implies: 
\begin{eqnarray}
\label{RGRequiredPot}
V(r) &=& {1\over r^4} \Bigl\{ c_4^{(1)} +  c_4^{(2)} \alpha(r_0)  \nonumber \\
&& \qquad \left. + \alpha^2(r_0)
\left[ c_4^{(3)} + B c_4^{(2)} \, \log\left( {r\over r_0} \right) \right]
+ \cdots \right\} \nonumber\\
&=& {1\over r^4} \Bigl\{ c_4^{(1)} +  c_4^{(2)} \alpha(r) 
 + c_4^{(3)} \alpha^2(r)  + \cdots\Bigr\} , 
\end{eqnarray}
where $\alpha(r)$ is the solution to the one-loop renormalization flow
$r d\alpha/dr = B \, \alpha^2$:
\begin{equation}
\label{RGRunning}
\alpha(r) = {\alpha_0 \over 1 - B \,\alpha_0 \, \log\left({r/\ell} \right)} .
\end{equation}
As usual, this expression is accurate to leading order in $\alpha_0$, but to all orders 
in $\alpha_0 \log(r/\ell)$. Standard calculations give $B = +3/2$ for real scalars, 
and $B = -1/2$ for complex scalars, in six dimensions. 

Several conclusions may be drawn from eq.~\pref{RGRequiredPot}. First, it shows
that the use of the renormalization group to resum all orders in $\alpha_0 \log(r/\ell)$
permits the inference of the coefficient of every power of $\log(r/\ell)$ 
to leading order in $\alpha_0 = \alpha(\ell)$. Furthermore, 
eq.~\pref{RGRequiredPot} establishes that this leading $\log(r)$ dependence 
is purely determined by the known one-loop renormalization group coefficient, 
$B$, and the two-loop vacuum energy coefficient, $c_4^{(2)}$, whose evaluation
is our next task.

Before turning to this task, there is another lesson to be drawn from 
eq.~\pref{RGRequiredPot} concerning the domain of validity of our conclusions. 
This equation shows that $V(r)$ depends logarithmically on $r$ only implicitly, 
due to its dependence on $\alpha(r)$. This implies the stationary point, $r_s$, 
of $V(r)$ occurs for $r_s$ satisfying the condition 
\begin{eqnarray}
\label{StationaryCondition}
-4 c^{(1)}_4 - 4 c^{(2)}_4 \alpha(r_s) + 
\alpha^2(r_s) \left[ B c^{(2)}_4 - 4 c^{(3)}_4 \right] + \cdots = 0 ,\nonumber
\end{eqnarray}
which is satisfied by $\alpha(r_s) = {\cal O}(1)$ (unless $\left| 
c^{(1)}_4\right| \ll \left| c^{(2)}_4 \right|$, in which case $\alpha(r_s)
\approx -c^{(1)}_4/c^{(2)}_4 \ll 1$). 

From these observations we see that 
\begin{equation}
\label{ronelleq}
{r_s \over \ell} = \exp\left[{1 \over B} \left( {1 \over \alpha(\ell)} - { 1 \over
\alpha(r_s)} \right) \right],
\end{equation}
and so $r_s\gg\ell$ follows from the logarithmic running of $\alpha$, given a 
reasonably modest difference between $\alpha(r_s)$ and $\alpha(\ell)$, provided
this running occurs within the perturbative regime $\alpha \lsim 1$. There
are two cases to consider: 
\begin{itemize}
\item
$B>0$ ({\it Real Scalars)}: For this choice $\alpha$ is asymptotically free, and
so $r_s > \ell$ requires $\alpha(r_s) > \alpha(\ell)$. In this case all of the
$c_4^{(l)}$'s -- and so also $\alpha(r_s)$ -- can be $O(1)$, and so a large
heirarchy is ensured for modestly small $\alpha(\ell)$. Although in this case
the precise {\it value} of $r_s$ cannot be computed in perturbation theory, 
its order of magnitude {\it is} known reliably to be of order 
$\exp[1/B\alpha(\ell)]$ compared to $\ell$.
\item
$B<0$ ({\it Complex Scalars}): In this case $\alpha(r)$ falls as $r$ increases,
so $r_s >\ell$ implies $\alpha(r_s) < \alpha(\ell)$. The use of the perturbative
running of $\alpha$ therefore requires $\alpha(r_s)\lsim 1$, and so
consistency requires $c_4^{(1)}/c_4^{(2)}$ to be small and negative. 
Remarkably, we find below that this condition is satisfied for some 
twistings of the scalar on a torus.
\end{itemize}

\vtop{
\includegraphics{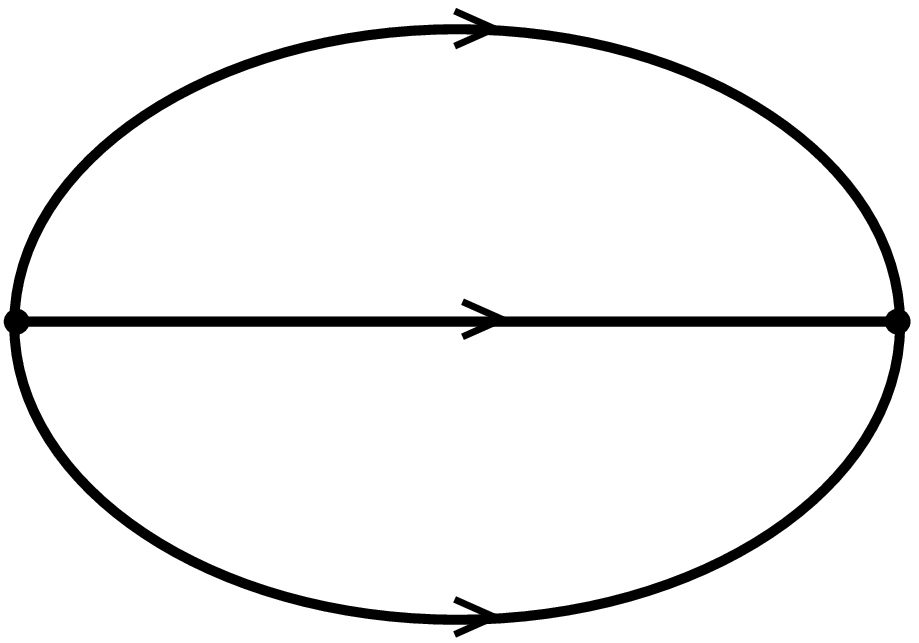}
\vskip 5.5cm
{\small Figure 1: The Feynman graph which gives the two-loop
contributions to the Casimir energy. To this must be added
the one-loop contribution into which the one-loop self-energy 
counterterm is inserted.}}
\bigskip

\noindent
{\it Calculating $c_4^{(2)}$:}
The two-loop correction to the vacuum energy arises from the two-loop vacuum graph
of Fig.~(1), where the arrows denote the direction of charge flow for complex scalars.
As computed in detail in Appendix B (for complex scalars), after renormalization this gives
the following finite contribution to the radion potential:
\begin{eqnarray}
\label{TwoLoopCase}
V_2(r) &=& {\alpha \over r^4} \; f_{ab}(\mu r) ,\nonumber \\
&\to& c_4^{(2)}(a,b) \; {\alpha \over r^4} \qquad (\mu \to 0),
\end{eqnarray}
where the function $f_{ab}(\mu r)$, which is symmetric in the twist numbers $(a,b)$,
is computed in Appendix B. Numerical integration of the obtained expressions
gives for the constant $c_4^{(2)}(a,b) = f_{ab}(0)$ the
values presented in Table (1). The result for real scalars is one half the 
$(0,0)$ entry for complex scalars.

Recall for complex scalars
(for which $B<0$) use of perturbation theory for the running of $\alpha$
requires $-\alpha(r_s) = c_4^{(1)}/c_4^{(2)}$ to be negative and small. 
This condition is satisfied for the two cases where there is a twist 
around only one cycle of the torus, with $\alpha(r_s) = 0.27$ for 
$(a,b)=(0,1/3), (1/3,0)$ and $\alpha(r_s) = 0.15$ for $(a,b)=(0,2/3),
(2/3,0)$. Using
$\alpha(\ell) = \infty$ and $B=-1/2$ in eq.~\pref{ronelleq} then gives
$r_s/\ell = 1.7 \times 10^3$ for $(a,b) = (0,1/3),(1/3,0)$ and $r_s/\ell =
8.3 \times 10^5$ for $(a,b) = (0,2/3),(2/3,0)$.

Although it is remarkable that some choices of complex-scalar boundary conditions 
satisfy the consistency conditions and generate large $r_s/\ell$, 
we emphasize that other choices for $(a,b)$ may also be consistent and
larger values of $r_s/\ell$ may be obtained, depending on the particle
content of the model. This is because the total value for $c_4^{(1)}$
is obtained by summing the contributions of all of the massless six-dimensional
states of the theory. In particular, a hierarchy as large as $r/\ell \sim 10^{15}$
(as would be required if $1/r \sim 10^{-3}$ eV) would require only
$\alpha(r_s) = 0.055$ if $\alpha(\ell) = 1$.

Notice, however, that the two-loop contribution comes
only from the scalars, and is positive for any nonzero twist, so any
model with twisted complex scalars must have its particle content arranged
to ensure $c_4^{(1)}$ is small and negative. Positive $c_4^{(1)}$ requires
the use of untwisted scalars (real or imaginary), for which $c_4^{(2)} < 0$.

To summarize, we see that a very light six-dimensional scalar field with
a cubic self-coupling term can naturally generate logarithmic 
radion potentials. For instance, in the case of a complex massless
scalar ($\mu = 0$) compactified on a torus, and satisfying twisted boundary 
conditions about one or more of the torus' cycles, the leading large-$r$ potential which 
is generated is given by eqs.~\pref{RGRequiredPot} and \pref{RGRunning}, with 
$B = -1/2$ and $c_4^{(1)}$ and $c_4^{(2)}$ are given by Table (1).
Notice that the leading coefficients of all powers of $\log r$
are determined by the running of the coupling $\alpha$, and are resummed by 
standard renormalization group arguments. The result for a real scalar
may be obtained by halving the result for an untwisted complex scalar.

Furthermore, this logarithmic potential naturally has the features identified 
in the introduction to generate an exponentially large hierarchy for $r/\ell$, since
higher powers of logarithms have systematically smaller coefficients. This is all
the more remarkable given that the only choices which can be made are the boundary
conditions which are satisfied by the scalar, and the massless particle content
which contributes to the one-loop Casimir energy. In particular, the relative signs
required for the existence of exponentially large stationary points for $V(r)$
need not have worked out as well as they did.

To this point, however, we are trading the puzzle of why $r/\ell$ is large
for the puzzle of why there should be a light six-dimensional scalar field
with cubic self-couplings.
We now turn to a search for microscopic models which might be expected to
have such scalars.

\section{Towards Microscopic Model Building}

In this section we ask what is required of a more microscopic model in order
to provide the desired light scalar, whose couplings generate the logarithmic
potential. Our purpose is twofold. We first intend to describe general features
any such model must have, and to identify the naturalness issues which any such
model must address. Although we examine several six-dimensional supergravity models in
detail, we do not succeed in obtaining a completely natural candidate.

\subsection{Model-Building Issues}

The framework for more microscopic model-building depends crucially on how
large are the radii which we are willing to contemplate. The two main options
divide over whether $r$ is larger or smaller than $(1 \,\TeV)^{-1}$. 
If $r \le (1 \,\TeV)^{-1}$, considerable latitude exists because there are fewer 
constraints on the model-building. The intermediate-scale string \cite{BIQK}
provides an attractive version of this scenario with $1/\ell \sim 10^{10}$ GeV,
$1/r \sim 1$ TeV and $1/(M_p r^2) \sim 10^{-3}$ eV. Here we will focus
on the more ambitious option, with $r \gg (1 \,\TeV)^{-1}$, with an eye to applications
for which $1/r \sim 10^{-3}$ eV. For radii this large, some general
observations are immediate.

For radii larger than the weak scale, model-building must take place within
the brane-world scenario, in which all (or most) standard-model particles
are confined to a four-dimensional surface within the larger six-dimensional
space \cite{ADD,RS}. (There may also be other particles confined to
other branes, with which ordinary particles only couple indirectly, 
the exchange of `bulk' states, which are free to move throughout the
six dimensions. There is some freedom to choose what kinds of particles
live in the bulk, but the bulk sector must include the graviton.
In what follows we assume this framework to be true, although for simplicity, 
and in keeping with our earlier calculations with
flat space, we imagine no cosmological constant in the six dimensions
(the ADD scenario \cite{ADD}), with gravitons not localized around the brane. 

In this case $1/r \sim 10^{-3}$ eV is as large a radius as can be contemplated,
due partly to the many bounds on modifications to Newton's Law on scales
larger than a millimetre \cite{ForceBounds}, and partly to limits on particle emission in
astrophysical environments like supernovae \cite{SNProbs,LED}. We believe models predicting
radii close to $10^{-3}$ eV are not yet ruled out in principle by these bounds, although
they must be rechecked once specific models are proposed. 

In any braneworld scenario with $1/r \sim 10^{-3}$ eV, there can be at most
six dimensions, and the scale of physics
on the branes themselves must be $M_b \sim 1$ TeV. This is because the four-dimensional
Planck's constant in these models is of order $M_p^2 \sim M_b^{2+n} r^n$, where $n$ is
the number of extra dimensions. For $n=2$ (six dimensions) this states that
$M_b \sim (M_p/r)^\half \sim 1$ TeV, as claimed. For $n > 2$, $M_b$ is unacceptably
low: {\it e.g.} when $n=3$, $M_b 
\sim (M_p^2/r^3)^\frac{1}{5} \sim 1$ GeV. (Of course it was this
argument run backwards which led early workers to contemplate radii as large as 
we are considering.)

We seek models having very light, six-dimensional scalars whose cubic self-couplings are not
systematically suppressed by powers of $1/r$.  The first of these conditions
is actually very easy to achieve, since the existence of very light six-dimensional
scalars arises very naturally within the braneworld framework. To be six-dimensional
the scalars must live in the bulk, and not be localized on the brane (as are ordinary
particles, like photons and electrons). Such scalars would be sufficiently light if
they were tied by a symmetry to another massless bulk particle, such as the graviton.

For instance, in any supersymmetric variant of the braneworld scenario (with millimetre-scale
extra dimensions) supersymmetry must be directly broken 
at the brane scale, $M_b \sim 1$ TeV, and so cannot be hidden too much from
ordinary matter in order to split supermultiplets by TeV scales. To the extent
that bulk states only couple to the branes with gravitational strength -- certainly
true for the graviton supermultiplet -- it follows that the mass splittings 
within the supermultiplets in the bulk are very small. They are small because 
they are suppressed by their weak gravitational couplings to the supersymmetry-breaking
sector, and so are of order $\Delta m \sim M_b^2/M_p$. 
Indeed, if $M_b \sim 1$ TeV this mass splitting is precisely the 
size of interest to us: $\Delta m \sim 10^{-3}$ eV. 

In this way we see that bulk scalars in general, and in particular
scalars that are tied to the graviton by supersymmetry, can be naturally
expected to have masses which are sufficiently small. They are sufficiently
light because the connection between $M_p$ and $r$ in six-dimensional models
ensures the coincidence of the scales $1/r$ and $M_b/M_p^2$. Furthermore,
the gravitational strength of their couplings would have hidden them 
from discovery before now. 

We are led to ask: do scalars arise in plausible representations of six-dimensional 
supergravity? The answer is `yes', including (in a particular sense, explained next) 
the graviton multiplet itself. Although strictly speaking the basic 
six-dimensional graviton supermultiplet consists of the sechsbein, gravitino and a 
Kalb-Ramond field \cite{MS,NS,SS}:
\begin{equation}
\label{6DGravityMultiplet}
\Bigl\{ {e^{\sss A}}_{\sss M}, \psi_{\sss M}, B_{\sss MN}^+ \Bigr\} ,
\end{equation}
this multiplet does not admit a Lorentz-invariant action because the
Kalb-Ramond field strength is self-dual. Consequently, six-dimensional supergravity
lagrangians are written with the supergravity multiplet coupled to a
Kalb-Ramond matter multiplet,
\begin{equation}
\label{6DKRMultiplet}
\Bigl\{ B_{\sss MN}^-, \chi, \sigma \Bigr\},
\end{equation}
which contains an anti-selfdual Kalb-Ramond field plus a fermion and scalar.
Thus the {\it nonchiral} gravity multiplet naturally contains a scalar, $\sigma$,
which plays the role of the dilaton in low-energy string compactifications.

Other spinless particles, which are tied by supersymmetry to other massless states in
the bulk -- like gauge bosons or chiral fermions -- are also candidates for our 
light scalar, although they introduce somewhat more model-dependence than does the dilaton.

So far, so good: we have a light, six-dimensional scalar. The thorny issue for making
models is obtaining a sizeable cubic self-interaction for these scalars. On one hand, the
explicit calculation of the light-scalar couplings is difficult to perform, since both 
the light scalar and the radion often have no potential energy at all unless supersymmetry
is spontaneously broken. Since our understanding of the nature of supersymmetry breaking
in brane models is poor, it is difficult to definitively decide how big the cubic
(and other) couplings might be in the low energy theory, once the brane physics is
integrated out.

On the other hand, it is very natural to assume that any cubic couplings obtained after 
supersymmetry breaking should be of the same order of magnitude as are the scalar masses
themselves. For instance, to the extent that the scalar potential for a light 
scalar, $\varphi$, is a supersymmetry-breaking effect, one might expect the entire potential
to have the schematic form:
\begin{equation}
\label{GenPotFormSS}
U(\varphi) = \left({M_b \over M_p}\right)^2 
\left[ {k_1 M_b^2 \over 2} \; \varphi^2 + {k_2\over 3!} 
\; \varphi^3 + {k_4 \over 4! \, M_b^2} \; \varphi^4 + \cdots \right],
\end{equation}
which expresses the suppression of all supersymmetry-breaking effects by $1/M_p^2$. 
(Here the quantities $k_i$ denote dimensionless $O(1)$ numbers.) 
Although such a potential naturally gives a scalar mass of order $\mu\sim M_b^2/M_p\sim 1/r$,
it also predicts a cubic coupling of order $g \sim (M_b/M_p)^2 \sim 1/(M_b r)^2$.
Unfortunately, such a coupling cannot give a purely logarithmic potential, since
it only contributes to higher order in $1/r$.
This is what happens in the simplest models of supersymmetry breaking, as we now
see. 

\subsection{Supergravity Models}

Consider a model consisting of the nonchiral supergravity multiplet, 
as above, together with an additional matter multiplet consisting of a gauge potential
and its superpartner: $\{A_{\sss M}, \lambda \}$. With this model we take the 
dilaton, $\sigma$, to be the six-dimensional light scalar whose cubic 
couplings are of interest. 

Rather than constructing its coupling to a brane, for calculational simplicity 
we instead break supersymmetry by artfully choosing the manifold on which we 
compactify the theory. In what follows we consider two such compactifications. We
first compactify on a sphere, and generate in this way a cubic scalar coupling. Although
this coupling is too small to generate a logarithmic potential, for reasons which are
much as indicated above, it is nonetheless instructive. Our second compactification is
on a torus, where we break supersymmetry completely using a 
Scherk-Schwarz mechanism \cite{SSM}. 
(That is, we break supersymmetry by assigning different boundary conditions to 
different members of a supermultiplet.) With an eye to obtaining
a positive radion potential, we require the dilaton to be antiperiodic about one of the
cycles of the torus, and keep all of the rest of the fields periodic about both cycles.
This compactification has the virtue of allowing many of the results of the previous
section to be carried over in whole cloth.

The bosonic part of the supergravity action, coupled to these matter multiplets,
is \cite{MS,NS}
\begin{eqnarray}
\label{6DSGBoseAction}
{{\cal L}\over \sqrt{\cal G}} &=& - \, {1 \over 2 \kappa^2} \; {\cal R} 
- \half \; \partial_{\sss M}\sigma \, \partial^{\sss M} \sigma 
- {1 \over 12} \; e^{2 \kappa \sigma} G_{\sss MNP} G^{\sss MNP} \nonumber\\
&& \quad - \, {1 \over 4}\; e^{\kappa \sigma}\, F_{\sss MN} F^{\sss MN} 
- {2 q^2 \over \kappa^4} \; e^{-\kappa\sigma}, 
\end{eqnarray}
where $G_{\sss MNP} = 3\partial_{\sss [M}B_{\sss NP]}+3 \kappa F_{\sss [MN}
A_{\sss P]}$ is the Kalb-Ramond field strength, $F_{\sss MN}$ is the usual
abelian field strength for $A_{\sss M}$, and all spinors carry a common
charge, $q$, under the gauge group.\footnote{Since $\psi_{\sss M}$ and $\lambda$ 
share the same chirality, while $\chi$ has opposite chirality, this theory as it stands
has anomalies. We imagine these to be cancelled either by the addition of more
matter fields or through a Green-Schwarz shift of $B_{\sss MN}$, without
affecting the rest of our analysis.} Recall that in six dimensions the couplings
have dimension $\kappa \propto \ell^2$ and $q \propto \ell$.

{\it Compactification on a Sphere:}
The equations of motion obtained from this action admit a solution consisting of
constant (and arbitrary) $\sigma$ ($\partial_{\sss M}\sigma = 0$), 
a gauge potential which is of the magnetic monopole form -- with monopole number 
$\pm 1$ -- in two dimensions, and a metric which is the product between flat 
space and a two-sphere having radius $r$ \cite{SS}: 
\begin{equation}
\label{MetricForm}
{\cal G}_{\sss MN} = \pmatrix{\eta_{\mu\nu} & 0 \cr 0 & r^2 \, \hat{g}_{mn} \cr} .
\end{equation}
$\hat{g}_{mn}$ is the metric on the (unit) two-sphere, and all other fields vanish:
$G_{\sss MNP} = \psi_{\sss M} = \chi = \lambda = 0$. This compactification is known
as the Salam-Sezgin model \cite{SS}. The flatness of the four-dimensional metric
is only possible when the monopole number is $n = \pm 1$, and this may be understood 
from the fact that the solution leaves one four-dimensional supersymmetry
unbroken only with this choice for the monopole number.

With this compactification a potential is generated for $r$ and $\sigma$ at
tree level, due to the nonzero background values which are taken by the two-dimensional
curvature scalar and electromagnetic field strength. The potential may be written
as follows \cite{Halliwell}:
\begin{equation}
\label{TreePotentialSS}
V(r,\sigma) = {2 q^2 e^{-\kappa \sigma} \over \kappa^3 r^2} \left[
1 - {\kappa^2 \, e^{\kappa\sigma} \over 4 q^2 r^2} \right]^2 .
\end{equation}
We see that the particular combination $X = e^{\kappa\sigma}/r^2$ appearing within the
brackets has developed a potential, along whose minimum the potential vanishes.
The combination of $Y = e^{-\kappa \sigma}/r^2$ appearing as a prefactor in 
eq.~\pref{TreePotentialSS} is then seen to  be a modulus of the compactification, 
parameterizing a flat direction along the bottom of this potential. 

In principle, this has the form we seek. The field $X$ is a six-dimensional
scalar whose mass is naturally of order $1/r$. It also has a cubic self coupling,
as measured by the third derivative of the potential in the $X$ direction, evaluated
at the minimum. Once the remaining supersymmetry is broken, lifting the potential's
degeneracy along the $Y$ direction, one might hope to generate logarithmic terms
in $r$ along the lines described in the previous sections. 

Unfortunately, the fly in the ointment is the size of the cubic coupling, which
we see is of order $1/r^2$. Although this model nicely illustrates the existence
of light scalars, and the generation of a potential for them, it does not furnish
an example of a loop-generated logarithmic potential.

{\it Toroidal Compactification:}
An alternative is to compactify on a torus (or an orbifold if we should
like to assign twisted boundary conditions to the scalar fields), 
which allows us to use the results of 
the previous section's calculations for the radion potential. Toroidal 
compactifications are possible for the model in the case $q=0$, in which case
a classical solution is given by arbitrary, constant $\sigma$ and a flat 
six-dimensional metric. 

From the four-dimensional perspective, this vacuum solution preserves $N=2$ supersymmetry 
and so no potential is generated for $\sigma$ or $r$ to any order in perturbation theory.
In the absence of supersymmetry breaking, this can be seen explicitly at 
one loop as being due to the cancellation of the
contributions of the various particles:
\begin{equation}
\label{OneLoopCancellation}
V_1^{\rm susy}(r) = \sum_n V^u_n(r) = 0,
\end{equation}
where $V^u_n(r)$ denotes the contribution due to the $n$'th field, and
the superscript `$u$' is a reminder that the calculation is performed
supersymmetrically, such as with all fields satisfying untwisted boundary
conditions. 

Suppose we now break the supersymmetry, by assigning twisted boundary conditions
only to some members of a supermultiplet, labelled by $\tilde n$. Then the above cancellation no longer obtains, leaving 
the one-loop result:
\begin{eqnarray}
\label{OneLoopNoCancellation}
V_1(r) &=& \sum_{\tilde n}V^t_{\tilde n}(r) + 
\sum_{n \ne \tilde n} V^u_n(r) \nonumber\\
&=& \sum_{\tilde n} \Bigl[
V^t_{\tilde n}(r) - V^u_{\tilde n}(r)\Bigr] ,
\end{eqnarray}
where the superscript `$t$' denotes the result computed using twisted
boundary conditions.
In some circumstances the light scalar field itself can be among the 
twisted fields which break supersymmetry, such as for 
compactifications on orbifolds having a $Z_3$ symmetry.

In this model of supersymmetry breaking, the radiative-corrections computed in
the previous section would directly apply, if there were only a nonzero cubic
$\sigma$ coupling. Although there is no potential for $\sigma$ in the model as it 
stands, one can be generated by loop effects. Since our chosen method for
supersymmetry breaking implies such a potential must vanish as $r \to \infty$,
this should lead once more to cubic couplings which are proportional to powers of $1/r$.

Although we do not have a model which circumvents this difficulty, we do see
reasons to be hopeful one could be constructed. The basic problem is to construct
a model for which the scalar mass is suppressed by a symmetry, but where its 
cubic self-interactions are not similarly suppressed. One line of model building
which this suggests is to tie the scalar to massless spin-one particles by having
it lie in a gauge multiplet, since gauge boson masses can be forbidden by
unbroken gauge symmetries without also precluding their having cubic self-interactions. 
(Scalar moduli in toroidal compactifications indeed typically do fall into 
four-dimensional gauge multiplets of $N=2$ 
supersymmetry \cite{ToroidalCompactification}.) In this case one might 
plausibly hope that supersymmetry protects the scalar mass more strongly than it
does the scalar's cubic couplings. 

\section{Phenomenological Consequences and Constraints}

What is generic about these models is their prediction of extremely 
light fundamental scalars which are gravitationally coupled. The two
fields which are generic to the models we consider are the radion, 
$r$, and the light six-dimensional scalar (such as the dilaton,
$\sigma$, in the supersymmetric examples just considered). 
We now turn to a discussion of the phenomenologically relevant properties
of these scalars, and of the physical signatures which follow from these.

\subsection{Masses and Couplings}

The first question is the size of the scalar mass. We now compute this for
the radion field, $r$, although similar considerations may also apply for the
six-dimensional scalar, depending on the model. Although we have discussed
the radion potential in some detail in previous sections, to determine 
its mass from this we must also compute the radion kinetic terms.
Since $r$ begins its life as part of 
the six-dimensional metric ({\it c.f.} eq.~\pref{MetricForm}) this kinetic energy 
may be read off (at tree level) from the six-dimensional Einstein-Hilbert action. 

A straightforward dimensional reduction of this action using the metric
\begin{equation}
\label{MetricAnsatz} 
{\cal G}_{\sss MN} = \pmatrix{ \hat{g}_{\mu\nu}(x) & 0 \cr 0 & \rho^2(x) \,
h_{mn}(y) \cr},
\end{equation}
with $\rho = r/\ell$, gives the result:
\begin{eqnarray}
\label{KineticTerms1}
{\cal L}_{\rm kin} &=& - {1\over 2\ell^4} \;\int d^2y \; 
\sqrt{{\cal G}} \; {\cal R} \\ &=& - {r^2 \over 2\ell^4} \; \sqrt{\hat{g}} 
\left[R(\hat{g}) - 2 \left({\partial r \over r} \right)^2 + 
{\ell^2 R(h) \over r^2} \right],
\end{eqnarray}
where we adopt the conventional normalization $\int d^2y \; \sqrt{h} = \ell^2$.

The Einstein-Hilbert term may be canonically normalized by rescaling $\hat{g}_{\mu\nu}
= \rho^{-2} g_{\mu\nu}$, giving:
\begin{equation}
\label{KineticTerms2}
{\cal L}_{\rm kin} = - {1 \over 2\ell^2} \; \sqrt{g} \left[R(g) + 
4 \left({\partial r \over r} \right)^2 + {\ell^4 R(h)\over r^4}  \right].
\end{equation}

{}From eq.~\pref{KineticTerms2} it is clear that the redefinition
$r = \ell e^{\ell \xi/2}$ puts the kinetic term for 
$\xi$ into canonical form. Adding this to an assumed logarithmic form
for the potential, eq.~\pref{LogPot}, we have the four-dimensional
radion-graviton dynamics relevant to cosmology given by:
\begin{eqnarray}
\label{XiDynamics}
{{\cal L} \over \sqrt{g}} &=& -\; {1 \over 2 \ell^2} 
R(g) -\; \half (\partial \xi)^2 - V(\xi), \\
V(\xi) &=& e^{-\lambda \ell \xi} 
\left[ a_0 + {2 a_1 \ell } \xi/2 + \cdots \right] + 
{\cal O}\left[e^{-q \omega \ell \xi}\right] ,\nonumber
\end{eqnarray}
with $\lambda = p/2 + 2$ (so $\lambda=4$ if $V(r) \propto
1/r^4$). 

We note here in passing that although an exponential potential,
$V(\xi) = V_0 \, e^{-\lambda \ell \xi}$, follows generically (at
tree level) from the assumption of a power-law potential, $V(r) \sim 1/r^p$,
the prediction $\lambda = (p+4)/2$ found above is not as robust. 
For instance, if the
scalar potential were to mix $\xi$ with another field -- such as happened 
when $r$ mixed with $\sigma$ in the supergravity potential, 
eq.~\pref{TreePotentialSS}, of the previous section -- then $\lambda = (p+4)/\sqrt{2}$.

To see how this comes about, consider an extreme example, where the potential
at very low energies ($\sim 10^{-3}$ eV)  
is a function only of one combination of $\xi$ and $N$ other canonically-normalized 
fields, $\varphi_i, i=1,...,N$: 
\begin{equation}
\label{NScalars}
V \propto \exp[-\lambda_0 (\xi + \varphi_1 + \varphi_2 + ... + \varphi_{\sss N})].
\end{equation}
In this case the canonically-normalized field which appears in the potential is
$\hat\xi = (\xi + \varphi_1 + \cdots + \varphi_{\sss N})/\sqrt{N+1}$ and so 
in terms of this variable the argument of the exponential is:
$- \lambda_0 \sqrt{N+1}\; \hat \xi $, leading to the prediction 
$\lambda = \lambda_0 \sqrt{N+1}$. The supergravity case has precisely
this form with $\lambda_0 = \frac{1}{2}(p+4)$ and $N=1$.

Regardless of the value found for $\lambda$, the mass which results 
from these manipulations is {\it extremely}
small, being of order 
\begin{equation}
\label{GenericMass}
m \sim {1 \over M_p r^2} \sim 10^{-33} \hbox{eV}
\end{equation}
if $1/r \sim 10^{-3}$ eV, with the decisive suppression by $M_p$ 
arising because of the radion's kinetic term sharing a common
origin with the four-dimensional Einstein-Hilbert action.

Before turning to the strong experimental constraints which 
any such scalar must satisfy, we make a brief aside to check
whether such a small scalar mass is technically natural.
That is, we ask if the small mass we have found is an artifact of the tree
approximation, or if it is stable under quantum corrections
and renormalization. 

\subsection{Naturalness} 

We now argue that masses as incredibly small as those of 
eq.~\pref{GenericMass} can be stable under radiative
correction, in models such as were considered in the previous sections \cite{NatSmall}. 
This remarkable stability may be seen most easily by integrating out 
physics at successive scales, and asking how large the contributions
to the scalar mass must be as these scales are integrated out. 

Consider first the contribution from energies much smaller than $1/r$. In
this energy range the effective theory is four-dimensional, and so the standard
analysis for a 4D scalar mass applies. Since the canonically normalized
radion, $\xi$, couples with strength $1/M_p$ to everything in the effective 4D 
theory, we expect loops to generically generate mass terms in the
potential which are of order $\delta V \lsim \Lambda^4 \, (\xi/M_p)^2$, 
with $\Lambda$ representing the largest energy scale relevant to the effective theory. 
Since use of a 4D theory presupposes $\Lambda \lsim 1/r$, 
we expect loop effects from these low scales to alter the radion mass by an 
amount $\delta m^2 \sim \Lambda^2/M_p \sim 1/(M_p \, r^2)$. Since these
are of the same order as the mass itself, such corrections do not destabilize
the radion mass.

Potentially more dangerous are quantum effects from scales in the range 
$1/r \lsim \Lambda \lsim M_b$. Naively one might imagine regarding the 
effective theory in this case as a complicated 4D theory, involving many
Kaluza-Klein states, and so again apply the 4D analysis just described. Using
$\Lambda \sim M_b \sim $ TeV would then lead one to the usual expectation
$\delta m \sim M_b^2/M_p \sim 1/r$, and so that scalar masses as small as 
$m \sim 1/(M_p r^2)$ must be fine-tuned. 

The 4D argument is misleading, however, because it hides many symmetries which
restrict the form which any UV-sensitive corrections must take. Specifically,
the effective theory for scales $\Lambda > 1/r$ is really six-dimensional, and
the radion -- being a component of the 6D metric -- is subject to the strong 
constraints of 6D locality and general covariance. The implications of these
conditions are hidden in a KK analysis because these conditions are difficult 
to enunciate in terms of the metric's KK modes.

In the 6D theory there are two kinds of UV-sensitive terms involving the 6D metric, 
${\cal G}_{MN}$, which can arise in the effective action. Those due to virtual bulk 
states may be written as local 6D effective interactions like
\begin{equation}
S_{\rm bulk} = \int d^6x \; \sqrt{{\cal G}} \Bigl[ A\Lambda^6 + B \Lambda^4 \, {\cal R} +
C \Lambda^2 \, {\cal R}^2 + \cdots \Bigr] , 
\end{equation}
where the ellipses represent other curvature squared terms, plus terms 
involving higher derivatives. $A,B$ and $C$ are dimensionless constants.
By contrast, those involving only virtual brane-bound states
are localized at the position of the branes, and have the form
\begin{equation}
S_{\rm brane} = \sum_b \int_{\Sigma_b} d^4y \; \sqrt\gamma \Bigl[ a_b
\Lambda^4 + \cdots \Bigr] ,
\end{equation}
where $\gamma_{\mu\nu}$ is the induced metric on brane $\Sigma_b$
and the omitted terms can
involve both extrinsic and intrinsic curvatures. $a$ is a dimensionless constant.

Suppose, first, that the extra-dimensional metric were roughly spherical, and so
${\cal R} \sim 1/r^2$. In this case $S_{\rm bulk}$ would imply a radion potential
of the form 
\begin{equation}
V(r) \sim A \Lambda^6 r^2 + B \Lambda^4 + C \Lambda^2/r^2 + \cdots \, .
\end{equation}
As stated earlier, like all other workers we put aside the cosmological constant
problem, and so take the six-dimensional cosmological term 
to vanish: $A\sim 0$. The vanishing of the $B$ term requires only that 
the extra dimensions have the topology of a torus.

(In passing we remark that within a supersymmetric theory a small bulk
cosmological constant might not be so hard to arrange. In the models of interest 
we have supersymmetry in the bulk space, with supersymmetry broken
at scale $\ell$ on the branes on which we live. Since all bulk-brane
couplings are gravitational in strength, we saw that supersymmetry-breaking
interactions of the four-dimensional gravity multiplet are suppressed
by powers of $1/M_p \propto 1/r$. On dimensional grounds, any 
supersymmetry-breaking mass splittings within the four-dimensional
gravity multiplet are at most of order $M_b^2/M_p \sim 1/r$, leading
one to expect the vacuum energy to be $A \Lambda^6 \sim 1/r^6$.)

Once the cosmological constant is removed, the quadratically-divergent 
curvature-squared term is seen to give a dangerous contribution to the radion
mass, since it predicts $\delta m \sim \Lambda/(M_p r) \gg m$. This contribution
does {\it not} arise for the toroidal geometries we are using, however,
because these are flat: ${\cal R} = 0$. The naturalness question in this language
then becomes the question whether quantum corrections can allow the extra-dimensional
curvature to satisfy ${\cal R} \ll 1/r^2$. 
We now argue that, for six-dimensional spaces, this can be so. 

In the absence of a bulk cosmological constant, the most UV-sensitive quantum 
contributions are to the brane tensions. Although the cosmological constant fine-tuning
requires $\sum_b a_b \Lambda^4 \sim 0$, it does not require the tension on
each brane to separately vanish. Furthermore, our knowledge of the spectrum of 
observable particles on our brane suggests $a_0 = O(1)$ for our brane at least. 
One might worry that large tensions like these must generate large 
extra-dimensional curvatures once they are considered as sources for the
gravitational field. 

We now argue that this does not happen. The key point is
that even a large $\Lambda^4$ type vacuum energy on the brane does not produce any
curvature in the extra dimensions. This is because in 6D this is NOT a 
cosmological constant, but is instead a delta function, localized at the
brane position. As a result, we must ask to what extent using this as a
source forces us into having a curved metric in the two extra dimensions.
It is a special property of 6D that point 3-brane sources do not induce
curvature in the transverse dimensions. This is because the gravitational
field of any straight system with co-dimension 2 (eg a cosmic string in 4D 
or a 3 brane in 6D) is strictly flat. The gravitational influence of such a
source arises because of the conical defect which is induced at the position of the
brane, rather than from local curvature. 

We see that the radion mass is protected from receiving large corrections
for two reasons. First, like all of the metric's KK modes, it must remain
massless if $1/r \to 0$ because in this limit it is tied to the massless
4D graviton by an unbroken symmetry: 6D Lorentz invariance. For flat
compactifications this suppresses the contributions from loops in the 
effective 6D theory -- {\it i.e.} for $\Lambda > 1/r$ -- to be of order
$\delta V(r) \sim 1/r^4$. The radion then acquires
a further mass suppression by $1/M_p$ because its kinetic term is {\it enhanced} by
$M_p^2$ relative to the potential. (The same is not true for other KK modes because
these acquire both their mass {\it and} their kinetic terms from the 6D Einstein action,
leading to no relative enhancement of the kinetic term relative to the mass term.)

\subsection{Kinetic Term Corrections}

Although they do not change the order of magnitude of the mass inferred,
radiative corrections to the kinetic terms do have an important impact
on our later confrontation with experimental bounds, and so we pause to
consider them briefly here.

We have seen that $\alpha(r)$ corrections can modify the potential 
at large $r$ in an important way, by adding logarithmic corrections in $r$.
The same is true for the radion kinetic terms, which are also dominated for
$r \gg \ell$ by logarithmic corrections arising due to powers of $\alpha(r)$.
Including these corrections, the relevant kinetic terms in the 
effective theory at large scales have the form:
\begin{equation}
\label{KineticTermsAgain}
{\cal L}_{\rm kin} = - {1 \over 2\ell^2} \; \sqrt{g} A^2[\alpha] 
\left[R(g) + 4 B^2[\alpha] \left({\partial r \over r}
\right)^2 \right],
\end{equation}
where both $A$ and $B$ have the series expansion $1 + O(\alpha)$. 

In general, the function $B$ changes the field redefinition, $\xi(r)$,
which is required to canonically normalize the radion kinetic terms. 
For general $A$ and $B$ we have 
\begin{equation}
\label{xidifference}
\xi = {2\over \ell} \int_\ell^r F[\alpha(r')] \; {dr' \over r'} ,
\end{equation}
where $F^2 = B^2 + \frac{3}{2} \, (A'/A)^2 \beta^2$, $A' = dA/d\alpha$
and $\beta = r d\alpha/dr$. 

$A$ and $B$ are independent of $r$ only to next-to-lowest order,
where $A \approx 1 + a \alpha_0$ and $B \approx 1 + b \alpha_0$,
and so to this order $\xi$ is still logarithmically related to $r$, 
but we have $\lambda \approx (\half p+2)[1 - b \alpha_0]$. 
Beyond this order
$\xi$ need not be strictly logarithmic in $r$, and so the
functional form of the potential $V(\xi)$ changes in
a more complicated way. We shall see that this has 
important implications for the cosmology of the radion field.

\subsection{Experimental Constraints}

We see there are three important mass scales for the bulk sector of these models:
the microscopic scale, $1/\ell$; the Kaluza-Klein mass scale, $1/r$; and
the radion mass scale, $1/(M_p r^2)$. These imply potentially interesting
modifications of gravity on a variety of scales.
For instance, the intermediate-scale scenario --- 
$1/\ell \sim 10^{10}$ GeV and $1/r \sim 1$ TeV --- puts $1/(M_p r^2)
\sim 10^{-3}$ eV into the millimetre range. Alternatively, for large
extra dimensions $1/r \sim 10^{-3}$ eV and $1/M_pr^2 \sim 10^{-33}$ eV.

\medskip
\begin{center}
{\it Millimetre Scales}
\end{center}
As the above examples show, very different choices for $\ell$ and $r$
imply modifications to gravity at scales of order $10^{-3}$ eV. 
The implications for experiments probing submillimetre range forces
can differ dramatically depending on the nature of the microscopic
physics. In the most extreme case $1/r \sim 10^{-3}$ eV and any observed
modifications to Newton's Law on these scales would signal the
transition to six-dimensional gravitational physics. Searches for such deviations
are now starting to probe forces within this interesting range \cite{Will}.

The exact kinds of signals experiments searching for these deviations
should expect to see depend on the precise nature of the couplings of the relevant
states to ordinary matter. Unfortunately these are 
difficult to cleanly predict, since they depend on the scenario involved. 
Furthermore, if the modifications are due to the onset of 6-dimensional
physics, then predictions are also hampered by
the large size of $\alpha(r)$ whose growth at low energies underlies our
mechanism for generating large radii. One generically 
expects deviations from the equivalence principle, and from the $1/r$ falloff
of the gravitational potential, at scales smaller than a millimetre.

\medskip
\begin{center}
{\it Very Long-Range Forces}
\end{center}
Since the masses of some of the lightest states, like the radion, 
can be incredibly small, $\sim 10^{-33}$ eV, their couplings are very strongly
constrained. Such small masses make the radion's Compton wavelength, $1/m$, 
of order $10^{26}$ m, permitting them to mediate extremely long-ranged
forces. We will argue here that the properties of the scalar predicted by
our mechanism for radius stabilization can evade these bounds, and do so
in an interesting way.

If the tree-level action of eqs.~\pref{KineticTerms1} and \pref{KineticTerms2}
were the whole story, our model would be ruled out. Since ordinary matter,
trapped as it is on a four-dimensional brane within the six dimensions of the
bulk, couples to the radion only indirectly through $\hat{g}_{\mu\nu}$, 
these expressions, with the field redefinition $\Phi = r^2$ show that the 
radion behaves precisely as a Brans-Dicke scalar, with coupling parameter 
$\omega = - \half$. Although Brans-Dicke scalars can be phenomenologically
acceptable, even if they are massless, solar-system tests require 
their couplings must be strongly suppressed relative to gravity, with current 
constraints requiring $\omega \gsim 3,000$ \cite{Will}.

The story is much more interesting once the radiative corrections due to $\alpha(r)$
are included, since these imply $\omega$ becomes a function of $r$. For weak
coupling, for instance, $\omega(\Phi) = -\half + \omega_1 \alpha_0 \log \Phi 
+ \cdots$. 

The recognition that $\omega$ is a function of $\Phi$ is crucial when comparing
with experiments, since it implies in particular that $\omega$ is a function of
time as the universe evolves cosmologically. Since the best limits only apply at
the current epoch, the constraints are satisfied so long as $\omega(t)$ approaches
sufficiently large values sufficiently quickly as the universe evolves towards
the present. Better yet, the evolution of $\omega(\Phi)$ in general scalar-tensor
theories has been found to generically be attracted towards large $\omega$
during cosmological evolution \cite{DamourNordtvedt}, indicating that current
bounds can be generically satisfied without making unreasonable
assumptions about the cosmological initial conditions. 

As we report in more detail in a companion publication \cite{ABRS}, 
we have applied the analysis of 
ref.~\cite{DamourNordtvedt} to realistic
cosmological evolution, using the functions $\omega(r)$ and $V(r)$ which
are plausibly obtained within our scenario. We find that current
constraints on post-Newtonian gravity and constraints from cosmology
can all be satisfied, and that it is also easy to arrange the
energy density of the scalar field to be currently starting to dominate the
energy density of the universe, making the radion a natural
microscopic realization of the `Planck scale quintessence' described
in \cite{AS1}.

\medskip
\begin{center}
{\it Accelerator Signals}
\end{center}

In the event that $1/\ell \sim $TeV and $1/r \sim 10^{-3}$ eV experimental
signals can be expected at future collider experiments, due to reactions
where bulk states are emitted and so appear as missing energy. 
These reaction rates are not $M_p$ suppressed
due to the cumulative contribution of the numerous KK levels of the
six-dimensional fields. Although it is beyond the scope of this paper to
provide detailed calculations of the signals expected for the emission 
of the six-dimensional scalars we consider here, these should add to
and closely resemble the signals which have been computed for the emission of 
six-dimensional metric states \cite{realgraviton,virtualgraviton} and six-dimensional
vector states \cite{LED}.

\section{Summary}

Our purpose has been to propose a mechanism for naturally generating
exponentially large radii, within a plausible scenario of extra-dimensional
physics. Besides making the usual assumption that there is no large
microscopic cosmological constant, our mechanism requires the following three ingredients:
\begin{enumerate}
\item
The universe must pass through a phase during which there are effectively
six large dimensions;
\item
A very light -- mass $m \lsim {\cal O}(1/r)$ -- spinless particle must be
present within the six-dimensional effective theory;
\item
The light scalar must have a cubic self interaction in the effective
six-dimensional theory, whose coupling $g$ is not itself already suppressed by powers
of $1/r$. 
\end{enumerate}
Under these circumstances we have shown that the renormalization of the one
marginal six-dimensional coupling, $g$, generically generates a potential
energy which depends logarithmically on the radius, $r$, of the 2 extra dimensions.

Besides generically having a runaway minimum, with $r \to \infty$, this potential
also has a minimum with $r/\ell$ exponentially large, where $\ell$ is the
length scale of the microscopic six-dimensional theory. The existence
of this minimum for perturbatively small couplings requires specific
relative signs between the contributions of successive loops in $V(r)$.
Within the context where the
radion potential is generated as a Casimir energy (such as due to the light 
six-dimensional scalar) we have shown that the required signs can be obtained.

In order to determine how natural our above three requirements are, we explored
several kinds of six-dimensional models explicitly. We have found that supersymmetry
can naturally assure items 1 and 2 in the list above, but assuring item 3 is more
difficult. In the models explored, supersymmetry both protects the scalar
mass {\it and} its cubic coupling to be of order $1/r$, violating item 3.
Unfortunately, it is difficult to exhaustively explore the potentials which
can be generated for such scalars given the present poor understanding of
how supersymmetry breaks. 

The same features of the radion potential which allow it to have naturally large
extrema, also allow the radion to evade all experimental constraints, 
despite the extremely small masses which are possible: $m \sim
10^{-33}$ eV. The evasion of these bounds arises through the cosmological 
evolution of the radion's effective couplings, and so  
rely crucially on the existence of the same radiative corrections
which generate the exponentially large radii. The role played by
these corrections is to convert the radion from an ordinary 
Brans-Dicke scalar to a more general scalar-tensor
theory for which the couplings to matter evolve cosmologically to small values.

If the leading contribution to the radion energy is positive (as may be arranged
by adjusting the model's particle content), the 
cosmological evolution of this very light radion suggests its interpretation 
as the source of quintessence, accounting for the present-day cosmological constant. 
We have verified that it is possible to construct a working quintessence 
model based on this picture, \cite{ABRS}, and believe that a more systematic
exploration of the cosmological implications of our radion-stabilization 
mechanism would be very fruitful.

\section{Acknowledgements}

We would like to acknowledge J. Maldacena, N. Seiberg, S. Shenker and
X. Tata for useful suggestions, and E. Poppitz for pointing out an error
in our treatment of our boundary conditions in an earlier version of this
work. We thank 
the Aspen Centre for Physics for providing the pleasant
environs in which this work started. CB's research is partially funded by NSERC 
(Canada), FCAR (Qu\'ebec) and the Ambrose Monell Foundation. AA and CS 
are supported by DOE grant DE-FG03-91ER40674 and U.C. Davis.

\section{Appendix A: The One Loop Casimir Energy}

In this appendix we provide expressions for the one-loop Casimir energy
(as a function of radius) of a complex scalar field $\phi$ with mass $\mu$
on the six-dimensional product of four-dimensional Minkowski space with a torus. 
The interaction Lagrangian ${\cal L}_{int} = (g/3!)(\phi^3 + c.c)$ is invariant under
$Z_3$ rotations $\phi \rightarrow w\phi$ where $w$ is a cubic root of unity, i.e.
$w^3 = 1$. We compute the Casimir energy for fields which satify the corresponding $Z_3$
invariant boundary conditions  about the torus' two nontrivial cycles. 
The momenta of the scalar in the toroidal directions are then 
$(n+a,m+b)$ in units of $2\pi/r$, where the constants $a$ and $b$ can take the values
0 for strictly periodic boundary conditions and 1/3 or 2/3 in the twisted cases.

The one loop vacuum energy for such a complex scalar, is given (for Euclidean momenta) by
\begin{eqnarray}
\label{1LCasimir}
\Lambda_1 &=& {1 \over r^2} \; \sum_{mn=-\infty}^\infty \int {d^4k \over (2 \pi)^4}
\; \log\Bigl(k^2 + \mu^2_{mn} \Bigr), \\
&=& - \, {1 \over r^2} \int_0^\infty {dy\over y} \int {d^4k \over (2\pi)^4} 
\sum_{mn = -\infty}^\infty \exp\Bigl[ -y(k^2 + \mu^2_{mn}) \Bigr],\nonumber
\end{eqnarray}
where
\begin{equation}
\label{MassLevels}
\mu^2_{mn} = \mu^2 + \left({ 2\pi \over r}\right)^2\Bigl[ (m+a)^2
+ (n+b)^2 \Bigr] .
\end{equation}

In the second of eqs.~\pref{1LCasimir} the integral over $k$ is now gaussian while the
sums can be performed explicitly in terms of Jacobi theta-functions. In particular we
will make use of
\begin{eqnarray}
\label{IntegralsSums}
       \theta_3(z,q) = \sum_{n = - \infty}^\infty q^{n^2} e^{2inz}
\end{eqnarray}
where customary $q = \exp(i\pi\tau)$. For the sums in  eqs.~\pref{1LCasimir} we then have
\begin{equation}
\sum_{n = - \infty}^\infty e^{-\pi x(n + a)^2} = e^{-\pi x a^2}\theta_3(i\pi a x,e^{-\pi x})
\label{BasicSum}
\end{equation}
with $x = 4 \pi y/r^2$. Using these in eq.~\pref{1LCasimir}, and integrating
the result over the torus to obtain a four-dimensional effective potential
\begin{equation}
\label{RhoLambdaDef}
V(r) = \int d^2z \; \Lambda = r^2 \, \Lambda,
\end{equation}
leads to 
\begin{eqnarray}
\label{OneLoopResultBare}
V_1(r) &=& -\; {1 \over r^4} \int_0^\infty {dx \over x^3}\;
e^{-\beta x}\;e^{-\pi x(a^2 + b^2)} \; \nonumber \\
&\times&  \; 
\theta_3(i\pi a x,e^{-\pi x}) \,\theta_3(i\pi b x,e^{-\pi x}) , 
\end{eqnarray}
where $\beta = \mu^2 r^2 /(4 \pi)$ and $a,b = 0,1/3$ or $2/3$ according to the boundary
conditions which are appropriate.

From the asymptotic forms for the theta functions 
\begin{equation}
\label{AsForm}
\theta_3(z,e^{-\pi x}) = 1 + \cos z\,e^{-\pi x}+ \cdots
\label{Largex}
\end{equation}
as $x \to \infty$ and
\begin{equation}
\theta_3(z,e^{-\pi x}) =  {1\over \sqrt{x}}\,e^{-\sin^2 z/\pi x}
\; \Bigl[1 + {\cal O}(e^{-\pi/x}) \Bigr], 
\label{Smallx}
\end{equation}
as $x \to 0$, we see that the one-loop vacuum energy converges in the infrared ($x \to \infty$) even if 
$\mu$ vanishes. For twisted boundary conditions this convergence is exponential,
reflecting the absence of exactly massless four-dimensional modes in this case.

On the other hand, the vacuum energy diverges in the ultraviolet 
($x \to 0$), but this divergence is independent of $r$
and so may be absorbed into a renormalization of the six-dimensional cosmological constant.
The finite, $r$-dependent result is obtained by subtracting the result for $r\to \infty$,
giving eq.~\pref{OneLoopResult}.

\section{Appendix B: The Two-Loop Casimir Energy}

In this appendix we compute the contribution to the Casimir energy coming from
the two-loop graph of Fig.~(1), plus the graph in which a wavefunction and mass
renormalization counterterm is inserted into the one-loop Casimir energy. 
We show that the result converges, as expected on general grounds, in the
infrared and ultraviolet, and evaluate the constant $c_4^{(2)}$ which controls
its overall size.

Our starting point is the contribution to the four-dimensional energy density
coming from the evaluation of Fig.~(1):
\begin{eqnarray}
\label{2LoopIntegral}
V_2^{\rm Fg2}(r) &=& - \; {g^2\over 6 \, r^2} \sum_{njkl} \int {d^4p\over (2\pi)^4}
\, {d^4q\over (2\pi)^4} \\
&&\quad \times  {1 \over (p^2 + \mu^2_{jk})(q^2 + \mu^2_{ln})
((p+q)^2 + \mu^2_{j+l,k+n})} .\nonumber
\end{eqnarray}
As before, $\mu^2_{nl} = \mu^2 + (2\pi/r)^2 [(n+a)^2+(l+b)^2]$, where
$a,b = 0,1/3,2/3$ reflect the scalar boundary conditions and $n,l$ are
integers. Use of the identity $1/X = \int^\infty_0 ds \; e^{-sX}$ for each propagator
permits the performance of the gaussian integrals. Using the basic result eq.~\pref{BasicSum},
we can then write the quadruple sum in eq.~\pref{2LoopIntegral} as
\begin{equation}
\label{HardSum}
\sum_{njkl} e^{-s\mu^2_{nj} - t\mu^2_{kl} - u\mu^2_{n+k,j+l}}
= e^{-\beta (s+t+u)} \; S_aS_b,
\end{equation}
where $\beta = \mu^2 r^2 /(4\pi)$ and we define the function $S_a$ by:
\begin{eqnarray}
\label{SDef}
S_a &=& e^{-\pi(u+t)a^2}\sum_{n= -\infty}^\infty e^{- \pi(u+s)(n+a)^2 -2\pi au(n+a)}\nonumber \\ 
&\times& \theta_3(i\pi(un+2au+at),e^{-\pi (u+t)}).
\end{eqnarray}
The function $S_b$ is defined correspondingly.

Combining everything gives the following contribution to $V(r)$:
\begin{eqnarray}
\label{Fig2Result}
V_2^{\rm Fg2} &=& - \; {g^2 \over 6r^4 (4\pi)^3}
\int_0^\infty {ds\, dt\, du\over (st + su + ut)^2} \nonumber \\
&\times& e^{-\beta(s+t+u)} \; S_aS_b.
\end{eqnarray}
Now the first question concerns divergences. The infrared limit corresponds to taking
$s,t,u$ large. For $u + t \gg 1$ we then obtain from the asymptotic form eq.~\pref{Largex} 
and the defining equation eq.~\pref{BasicSum} 
\begin{equation}
\label{LargeAss}
S_a =  e^{-\pi(s+t+4u)a^2}\theta_3(i\pi a(s+2u),e^{-\pi(s+u)}).
\end{equation}
Since $S_a$ thus goes at worst to unity in this limit, we see that the integral
converges in the infrared, even if $\mu \to 0$. 

It has several sources of diverence in the ultraviolet. The divergence when all three variables,
$s,t$ and $u$, vanish is removed by subtracting the result for $r\to\infty$,
implying that it is removed by renormalizing the six-dimensional 
cosmological constant. Using the asymptotic form eq.~\pref{Smallx}
for $x\to 0$, one finds for $u+t \ll 1$
\begin{equation}
\label{SAssForm}
S_a = {e^{-\pi wa^2}\over \sqrt{u + t}} \; \theta_3(i\pi aw,e^{-\pi w})
\Bigl[1 + {\cal O}\Bigl(e^{-\pi/(t+u)}\Bigr)\Bigr].
\end{equation}
where $w = (su + st + tu)/(u + t)$. The result, after subtracting the 
large-$r$ limit, is therefore obtained
from eq.~\pref{Fig2Result} by the replacement 
\begin{equation}
\label{HatSDef}
{S_aS_b} \rightarrow S_aS_b - {1 \over st+su+tu} .
\end{equation}

Although this subtraction renders finite the limit where $s,t,u$ all vanish
with fixed nonzero ratios, it does not cure the ultraviolet divergence when
two of the variables $s,t,u$ vanish with the third held fixed. This divergence
cancels with the result obtained when the counterterms for wavefunction and mass
renormalization are inserted into the one-loop Casimir energy. 

To see how this works, notice that an evaluation of the one-loop 
scalar self-energy for the torus is
\begin{eqnarray}
\label{OneLoopSE}
\Pi(p^2) &=& {g^2\over r^2} \int_0^\infty dt \, du \\
&\times& \sum_{ln}
\int {d^4k\over (2\pi)^4} \; e^{-t (k^2+\mu_{ln}^2) - u((k-p)^2+\mu_{ln}^2)} \nonumber
\end{eqnarray}
The integrals and sums here are of the same kind as before and we obtain
\begin{eqnarray}
\Pi(p^2) &=& {g^2 \over (4\pi)^2 r^2}  \int_0^\infty {du \, dt \over (u+t)^2} \; e^{-\beta (u+t)}
 e^{-\left( {ut\over u+t}\right) {p^2 r^2\over 4\pi} }\nonumber \\
&\times &\;e^{-x(a^2 + b^2)}\theta_3(ixa,e^{-x})\,\theta_3(ixb,e^{-x})
\end{eqnarray}
where $x = u+t$.
For an on-shell renormalization scheme  the mass and wavefunction counterterms are now found
by expanding this in powers of $p^2$. The resulting integrals are dominated by the contributions
from the region where $u + t \ll 1$ and gives:
\begin{eqnarray}
\label{DeltaZ}
\delta Z &=& {g^2 \over (4\pi)^3} \int_\epsilon^\infty du \, dt \; {ut \over (u+t)^4}
\; e^{-\beta(u+t)} \\
\mu^2 \delta Z + \delta \mu^2 &=& - \, {g^2 \over (4\pi)^2 r^2}
\int_\epsilon^\infty {du \, dt \over (u+t)^3} \; e^{-\beta(u+t)} 
\end{eqnarray}
where we eventually will take the regulator $\epsilon \rightarrow 0$ in the ultraviolet limit.
Denoting $\xi(p^2) = \delta Z (p^2 + \mu^2) + \delta \mu^2$, the insertion
of these counterterms into the one-loop Casimir energy gives
\begin{eqnarray}
\label{OneLoopCTResult}
&&V_2^{ct}=  - \, {1 \over 2r^2} \int_0^\infty ds 
\sum_{ln} \int {d^4p\over (2\pi)^4} \; \xi(p^2) e^{-s(p^2 + \mu^2_{ln})} 
\nonumber\\
&=& {g^2 \over 6r^4(4\pi)^3} \int_0^\infty\!\! ds \, dt \, du 
\; e^{-\beta(s+t+u)}\, T_{ab}(s,t,u)
\end{eqnarray}
where we have symmetrized the result with respect to permutations of $s,t,u$
and the function $T_{ab}$ is defined by
\begin{eqnarray}
\label{Tdef}
T_{ab} &:=& \left[1 - {2 u t \over s (u+t)} \right] \; {1\over s^2 (t+u)^3} \nonumber \\
&\times& \;\left[e^{-\pi s(a^2 + b^2)}\theta_3(i\pi sa,e^{-\pi s})\,
\theta_3(i\pi sb,e^{-\pi s}) -{1\over s}\right] \nonumber \\
&+&  \hbox{cyclic permutations of $s,t,u$}. 
\end{eqnarray}

The two-loop contribution to the radion potential is obtained by summing 
$V_2^{ct}(r)$ with the result, eq.~\pref{Fig2Result}, from Fig.~(1). It is both ultraviolet
and infrared finite, and when evaluated at $\mu =0$ gives expression \pref{TwoLoopCase}
of the text.

\end{document}